\documentclass[aps,pre,twocolumn,showpacs,superscriptaddress,groupedaddress]{revtex4-2}
\usepackage{amsmath}
\usepackage{amssymb}
\usepackage{amsfonts}
\usepackage{graphicx}
\usepackage{dcolumn}
\usepackage{bm}
\usepackage{sidecap}
\usepackage{subfloat}
\usepackage{parskip}
\usepackage{grffile}
\usepackage{color}
\usepackage{hyperref}
\usepackage{hhline}
\usepackage{mathtools}
\usepackage{graphics}
\usepackage{multirow}
\usepackage{verbatim}
\usepackage{longtable}
\usepackage{rotating}
\usepackage{setspace}
\usepackage{epsfig}
\usepackage{subfigure}
\usepackage{epstopdf}
\usepackage{gensymb}
\usepackage[normalem]{ulem}
\usepackage{tikz}

\makeatletter
\def\@eqnnum{{\normalsize \normalcolor (\theequation)}}
\makeatother

\graphicspath{{./}{ER/main/}}

\begin{document}
	
	\title{Eigenvalue ratio statistics  of complex networks: Disorder vs. Randomness}
	\author{Ankit Mishra, Tanu Raghav and Sarika Jalan}
	\affiliation{Complex Systems Lab, Department of Physics, Indian Institute of Technology Indore, Khandwa Road, Simrol, Indore-453552, India}

	\begin{abstract}
		The distribution of the ratios of consecutive eigenvalue spacings of random matrices has emerged as an important tool to study spectral properties of many-body systems. This article numerically investigates the eigenvalue ratios distribution of various model networks, namely, small-world, Erd\H{o}s-R\'{e}nyi random, and (dis)assortative random having a diagonal disorder  in the  corresponding adjacency matrices. Without any diagonal disorder, the eigenvalues ratio distribution of these model networks depict Gaussian orthogonal ensemble (GOE) statistics. Upon adding diagonal disorder, there exists a gradual transition from the GOE to Poisson statistics depending upon the strength of the disorder. The critical disorder ($w_c$) required to procure the Poisson statistics increases with the randomness in the network architecture. We relate $w_{c}$ with the  time taken by maximum  entropy random walker to reach the steady-state. These analyses  will be helpful to understand the role of eigenvalues other than the principal one for various network dynamics such as transient behaviour.

	\end{abstract}

	\maketitle
	
	\section{Introduction}
	
	It was during the 1950s when E. Wigner envisioned that the complex Hermitian operator of heavy nuclei could be replaced by random matrices whose elements are chosen randomly from some distribution \cite{mehta}.  He and others proposed that the statistical properties of the eigenvalue spectrum of random matrices should mimic the original system under consideration without any detailed knowledge of the structure and show universality with the appropriate symmetry class of the system. There exist exactly three symmetry classes;  Gaussian orthogonal ensemble (GOE) for real, Gaussian unitary ensemble (GUE) for complex , and Gaussian symplectic
		ensemble (GSE) for quaternionic random numbers. The GOE  matrices remain invariant under orthogonal transformation, i.e., $A \rightarrow Q^{-1}AQ $ for any orthogonal matrix $Q$. Correspondingly, the other two symmetry classes remain invariant under unitary and symplectic transformations, respectively.
	Random matrix theory (RMT) found its application in different areas of research; statistical physics \cite{RMT_stats}, quantum chaos \cite{Rmt_chaos}, condensed matter physics \cite{RMT_locn}. For example, in tight binding models, it is used to characterize localized and delocalized states \cite{RMT_locn}. In quantum systems, RMT is often used to identify if the system is integrable, chaotic, or a mixture of both of them \cite{Rmt_chaos}.
	
	In the RMT framework, one usually compares the spectral fluctuation of the system with those predicted by RMT \cite{Dyson}. The nearest neighbor level spacings, defined as the difference between the consecutive eigenvalues of the given operator, is the most accepted spectral measure. However, to compare the spectral fluctuations, one needs to unfold the original eigenvalues to separate the smooth global part and fluctuating local part (system dependent), and then spacings are calculated on the unfolded eigenvalues \cite{brody}. Usually, unfolding procedures are not unique and non-trivial, which can lead to misleading statistical results \cite{unfolding_error1,*normal_mode_random}. For instance, in the case of the Bose-Hubbard model at considerable interaction strength, the density of states is not a smooth function of energy, and it becomes non-trivial to separate them into the smooth global part and fluctuating local part \cite{Unfolding_Bose_Hubbard}. Oganesyan and Huse solved this impediment of unfolding by introducing a new measure called as the ratio of consecutive eigenvalue spacings ($r$), independent of the local density of states and hence requiring no unfolding \cite{eig_ratio_first}. Additionally, it is easy to compute it with a lower computational cost.
	
	Furthermore, network science has witnessed tremendous growth in the last two decades due to its ability to comprehend and predict complex behaviors of many large scale real-world systems spanning from technology to social systems \cite{Rev_network,*camel_review,*SN_Dor,*newman_siam}. A network consists of nodes corresponding to the elements of a system and links representing interactions between these elements. To capture various  properties of real-world complex systems, different network models have been proposed. Among these Erd\H{o}s-R\'enyi (ER) random, scale-free, and small-world network models are the most popular ones \cite{Rev_network}.
	The ER random network model, theoretically investigated by Erd\H{o}s and R\'{e}nyi  in late 1950’s depict many fascinating phenomena including transition to the formation of a giant cluster with an increase in the probability of connecting nodes \cite{ER}. Despite the tremendous theoretical success of the ER random network model, it was  not  considered to imitate real-world networks due to its limitation to capture  properties  like high clustering and power-law degree distribution abundantly found in a diverse range of real-world systems. Watts and Strogatz in their landmark paper \cite{small-world} introduced the small-world network model which generates networks having very high clustering coefficient,  as that of regular lattice, and average shortest path length, as that of the  ER random networks, two properties readily witnessed in networks representing many real-world complex systems \cite{small_expt,*brain,*food_web}. It is important to note here that both the ER random and small-world networks have infinite dimension ($d \rightarrow \infty$), as the average shortest path length ($l$) scales as $l \sim ln(N)$ and $d$ can be calculated as $l \sim N^{1/d}$.
	
	Moreover, RMT has been extensively used in network science to capture phase transition and study various phenomenon. For example, using spectral statistics, localization transition was studied for ER random network, Cayley tree, and Barabasi-Albert scale-free networks \cite{LT_spec_stats}. The value of critical disorder as a function of average degree for Anderson transition was calculated for these model networks using the distribution of eigenvalue spacings. Further, to obtain a clear Anderson transition, a low value of the average degree as a threshold was proposed, above which no clear Anderson transition could occur for any of these networks. However, the absence of a transition could also be attributed to the small size of the considered networks. In Ref.~\cite{wave_locn}, using the level statistics, it was shown that Anderson-like transition can be obtained in complex networks without a diagonal disorder and just by tuning the clustering coefficient.  In Ref.~\cite{knight_epl}, RMT was applied on random geometric graphs (RGG), and it was found that as a deterministic connection parameter increases, eigenvalue spacing shows a gradual transition from the Poisson to the GOE statistics. Also,  spectral analyses have been carried  out for  random networks with an expected degree and $\beta$-skeleton graphs \cite{random_expectd,beta_sk}. Further, Ref.~\cite{uni_spars_rand} has shown the universality of nearest neighbors spacing distribution $(P(s))$ with network size by calculating the Brody parameter for random networks. However, it did not investigate impact of variation in the strength of diagonal disorder on spectral properties of complex networks.

	This paper investigates the statistics of the distribution of consecutive eigenvalue spacing ratios of adjacency matrix of various model networks having diagonal disorder,  and its interplay with the randomness caused by the occurrence of ones in off-diagonal elements representing pair-wise connections.
	The idea of adding the diagonal disorder was first originated in Anderson’s seminal paper for a tight-binding model having only nearest neighbors hopping \cite{ands_locn}. In Anderson's model, the diagonal disorder in the Hamiltonian matrices depicts the onsite potential of different sites. Diagonal disorders in complex networks, i.e  self-loops in the graph representation, may represent  various intrinsic properties of the nodes, and depending upon the system under consideration, they may carry different physical meanings. For example, in the case of excitatory dynamics of photosynthesis molecules, the diagonal disorder corresponds to the excitation energies of pigment molecules in different protein environment or imperfect fabrication of the structures \cite{dig_diso_1,*dig_diso_2,*dig_diso_4}.
		In an economic model with nodes representing firms and links representing interactions between firms in terms of their production, the diagonal disorder corresponds to the productivity of each firm \cite{Econ_1}.
		Another example is of optical systems where diagonal disorder is akin to variations in the refractive indices of the optical fibers, and connections represent random position of the fibers \cite{dig_diso_5,*dig_diso_6,*dig_diso_7,*dig_diso_8}.
	
	Further, note that the term 'distribution of the ratios of consecutive eigenvalue spacing' and 'eigenvalue ratio statistics' will mean the same in this paper and be used interchangeably. The eigenvalue ratio follows GOE statistics for all the model networks, namely small-world, ER random, and dis(assortative) ER networks. We show that upon  increasing the strength of the diagonal disorder in the adjacency matrix of a network, the eigenvalue ratio statistics gradually depicts a transition from the GOE to the Poisson statistics. However, for small-world networks, the critical disorder needed to obtain the Poisson statistics increases with an increase in the value  of the rewiring probability.
	Additionally, we probe the impact of degree-degree correlation or (dis)assortative degree mixing of networks on eigenvalue ratio statistics. We further relate the critical disorder required to obtain the Poisson statistics with the transient dynamics of a random walker.
	
	The paper is organized as follows. Sec.~\ref{a1} consists of definitions of eigenvalue spacing ratio, construction  of model networks, and measure of localization. Secs.~\ref{a3} and ~\ref{a4} contain results about effect of the interplay of disorder and randomness on the eigenvalue ratio statistics for various model networks. Sec.~\ref{a5} relates the critical disorder which is required to obtain the GOE to Poisson transition with the dynamics of maximally entropy random walker. Finally, Sec.~\ref{a6} concludes the article.

	\section{Methods and Techniques} \label{a1}

	A network denoted  by {\it G} = \{V,E\} consists of set of {\it nodes} and  {\bf \it links}. The set of  {\it nodes} is represented by $V = \{v_{1}, v_{2}, v_{3}, \ldots, v_{N}\}$ and 
	{\it links}
	with $E = \{e_{1}, e_{2}, e_{3}, \ldots ,e_{M}\}$ where  $N$  and  $M$  are sizes of $V$ and $E$ respectively.\ Mathematically, a network can be represented by its adjacency matrix $A$ whose elements 
	are defined
	as  $A_{ij} = 1$ if node $i$ and $j$ are connected and $0$ otherwise.\ 
	The eigenvalues of the adjacency matrix $A$ are denoted by $\left\{\lambda_{1}, \lambda_{2}, \lambda_{3},\ldots,\lambda_{N}\right\}$ where
	$\lambda_{1} \geq \lambda_{2} \geq $\ldots$ \geq \lambda_{N}$.

	We perturbed the adjacency matrix ($A$) by adding a diagonal matrix. The new adjacency matrix becomes $A' = A+D$, where $D$ is a diagonal matrix added to the original adjacency matrix. The diagonal elements of $D$ i.e. $D_{ii}$ are random numbers drawn uniformly from a box distribution between $(-w,w)$ with a width $2w$. We probe the effect of impact of an increase in the diagonal strength ($2w$) on the eigenvalue ratio statistics. 
	The eigenvalue ratio statistics is known to be very useful to identify localized and delocalized eigenvectors. Localization of eigenvectors means that few eigenvector entries take very high values compared with the other entries. On the other hand, in the case of delocalized eigenvector, all the entries take almost a equal value. Further, the eigenvalue ratio statistics corresponding to the localized eigenvectors is known to depict the Poisson statistics, while for delocalized eigenvectors, it is known to manifest the GOE statistics \cite{locn_stats}.
	
	\paragraph {Ratio of eigenvalue spacing:} 
	Following the Ref.~\cite{eig_ratio_first}, the  ratio of consecutive eigenvalue spacing is defined here as
	\begin{equation}
	r_{i} = \frac{min(s_{i+1}, s_{i})} {max(s_{i+1},s_{i})}
	\end{equation}
	where $s_{i}$ = $\lambda_{i+1}$$-$$\lambda_{i}$ is the spacing between eigenvalues $\lambda_{i+1}$ and $\lambda_{i}$ with  $i \in$ ($1,2,3 \hdots  N-1$). 
	Also, one can verify that $0 < r_{i}< 1$.
	Ref.~\cite{eig_ratio_first} provides
	only numerical estimation of $P(r)$ for the GOE distribution, and discussions on other two symmetry classes (GUE and GSE) were lacking. This gap was filled in Ref.~\cite{distribution_Ratio} where an exact distribution function of $r$ was derived which was not restricted only to GOE but includes GUE and GSE classes as well.
	The distribution function ($P(r)$) approximating GOE statistics is given as
	\begin{equation}
	\label{eq:pr_goe}
	P(r)  \sim \frac{54/8 \times (r+r^{2})}{(1+r+r^{2})^{5/2}}
	\end{equation} 
	and the theoretical average value of $r$ for GOE and Poisson statistics has been estimated to be equal to 0.53 and 0.38, respectively, with the distribution function given by 
	\begin{equation}
	\label{eq:poi}
	P(r) \sim \frac{2}{(1+r)^{2}}
	\end{equation}
	
	Note that there also exists an empirical formula  to capture the GOE to the Poisson crossover by scanning a fitting parameter \cite{Ratio_formula}. However, the present study uses only $\langle r \rangle$ to record the GOE to Poisson transition instead of any fitting parameter. Nevertheless, $\langle r \rangle$ is a well admissible parameter to capture GOE to Poisson crossover and widely used in different systems \cite{R_J_1,*R_J_4,*R_J_3,*R_J_5}. 
	
	\begin{figure}[t]
		\centering
		\includegraphics[width= 0.40\textwidth]{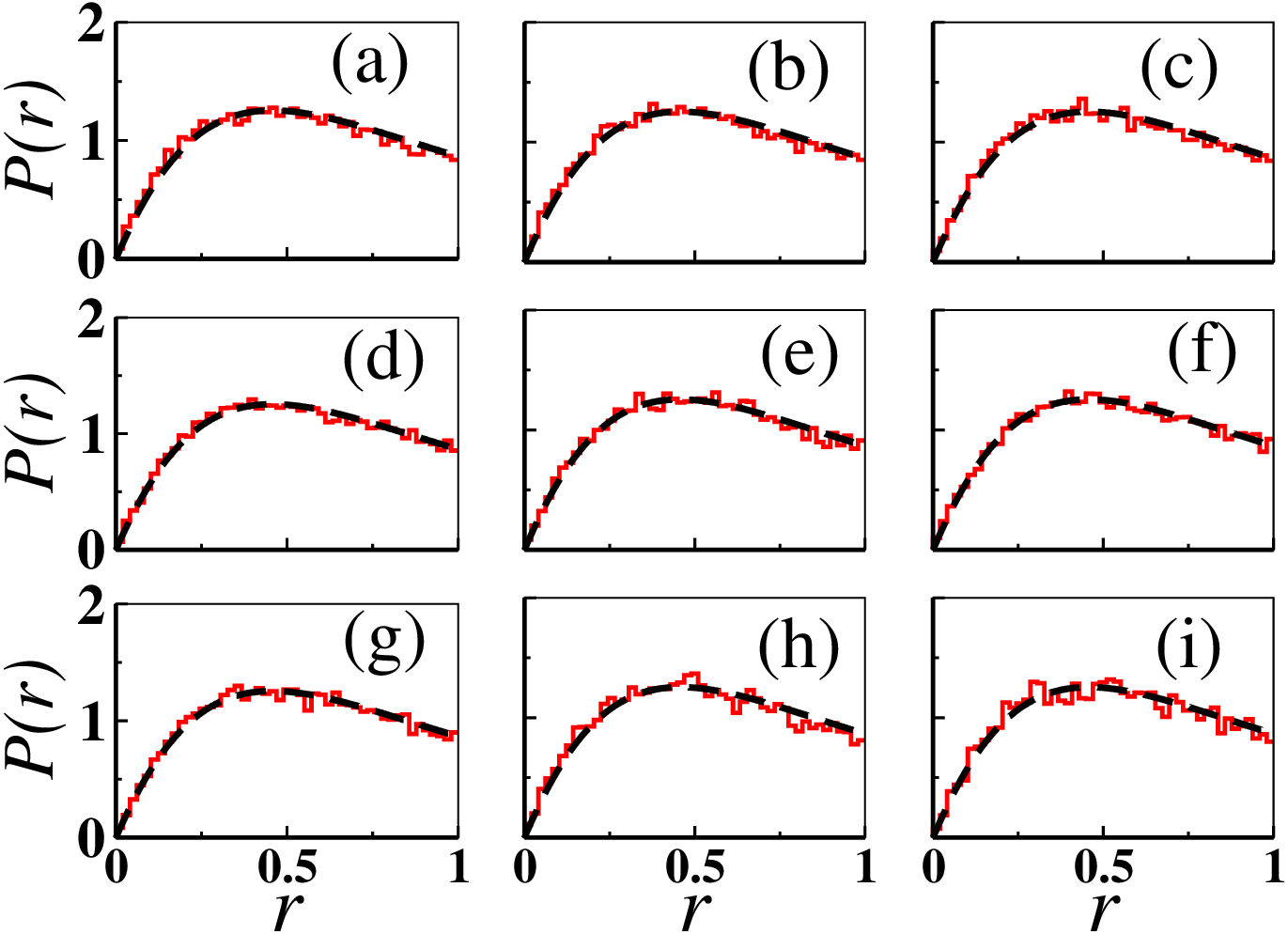} 
		\caption{(Color online) Distribution of ratio of consecutive eigenvalue spacing for various rewiring probabilities. 
			Black line (dashed) indicates distribution function for GOE statistics and red stairs indicate the data.
			Here, we have considered $N = 2000$ and  $\langle k \rangle = 20$ with $20$ network realizations.
			(a) $p_{r}$ = 0.001 (b) $p_{r}$ = 0.002 (c) $p_{r}$ = 0.005 (d) $p_{r}$ = 0.01 (e) $p_{r}$ = 0.02 (f) $p_{r}$ = 0.05 (g) $p_{r}$ = 0.1 (h) $p_{r}$ = 0.5 (i) $p_{r}$ = 1.}
		\label{wzr}
	\end{figure}
	
	\section{Results} \label{a2}
	\subsection{Eigenvalue ratio statistics of Small-World Networks}   \label{a3}
	We first analyze the eigenvalue ratio statistics for the small-world networks generated using the Watts and Strogatz algorithm. 
	    Starting with a 1D lattice, links are rewired with a probability $p_{r}$ such that 
		$0$ $\leq$ $p_{r}$ $\leq$ $1$. For some intermediate rewiring probabilities, the network undergoes the small-world transition characterized by a high clustering and a shorter average path-length. We numerically diagonalize the adjacency matrix to obtain its eigenvalues. We focus on the eigenvalues on the central part of the spectrum more precisely, inside the width $d\lambda \approx 1.5$ on both sides of $\lambda \approx 0$, a usual practice while analyzing the eigenvalues statistics to localization transition \cite{lam_0}. We wish to emphasize here that a slight increase or decrease in the width does not affect the results which is discussed later. It is evident from Fig.~\ref{wzr} that $P(r)$ fits very well with the exact form of the GOE statistics (Eq.~\eqref{eq:pr_goe}) for all the values of the rewiring probability. We then calculate the average value of $r$  numerically using Simpson's rule, which comes out to be around $0.52$ for all $p_r$ values  and thus validate the GOE statistics.
	We would also like to mention here that a similar observation was found through the distribution of eigenvalue spacings in Ref.~\cite{JS}. The authors had shown a change in the Brody parameter ($\beta$) value with the rewiring probability, finally leading to the GOE transition at the onset of the small-world transition.\ However, in \cite{JS}, authors had considered the entire eigenvalue spectrum to depict the GOE transition, whereas the present study focuses only on the central part of the eigenvalue spectrum which shows the GOE statistics for all  $p_r$ values.\\

\begin{figure}[t]
		\centering
		\includegraphics[width=0.48 \textwidth]{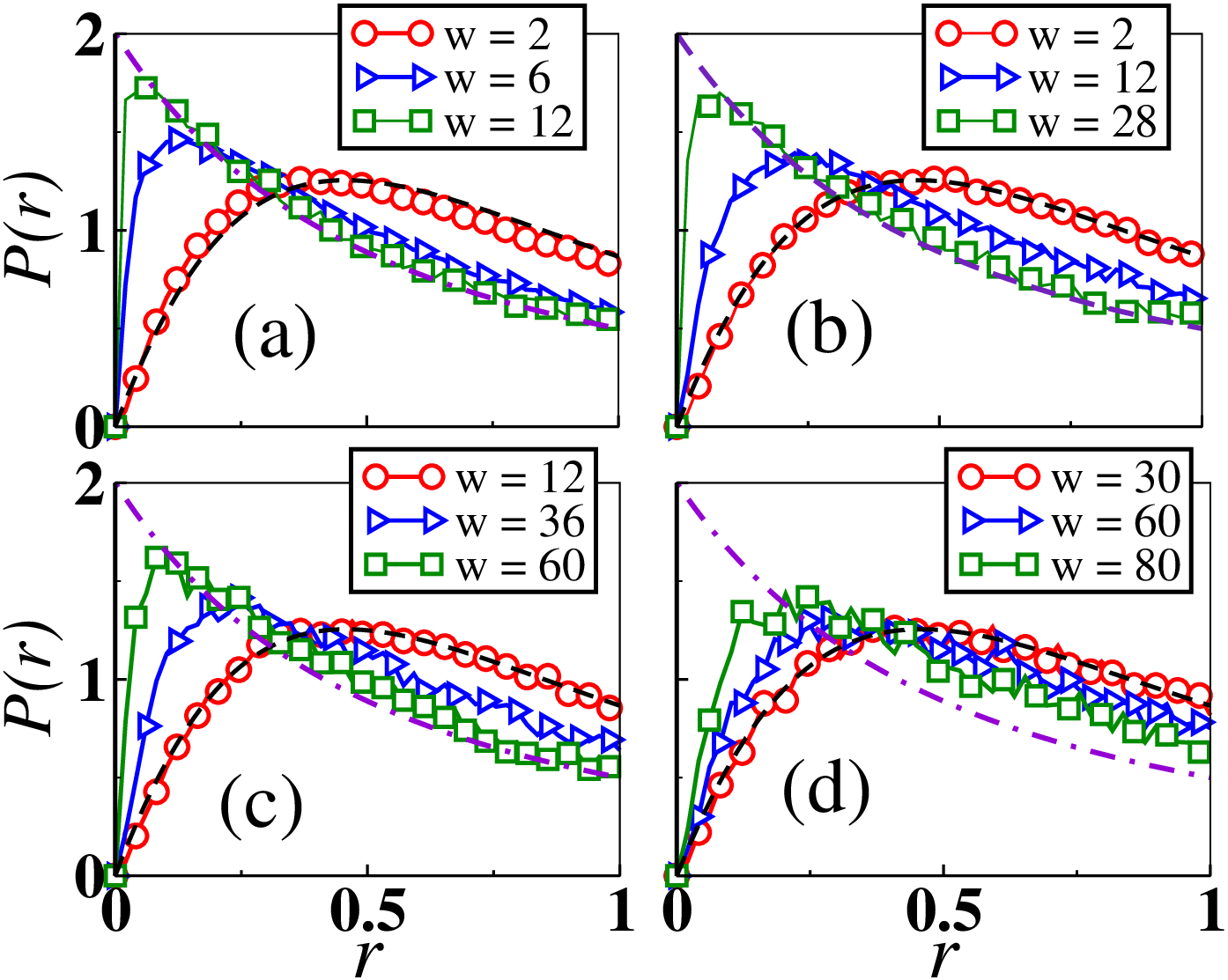}\caption{(Color online) Distribution of ratio of consecutive eigenvalue spacing of adjacency matrices having diagonal disorder drawn from uniform distribution of width $2w$ for different values of $p_r$. 
			Black (dashed) and violet (dashed-dotted) lines, respectively indicate GOE and Poisson distribution function.
			Here, $N = 2000$ and  $\langle k \rangle = 20$ with $20$ network realizations and $40$ disorder realizations. (a) $p_{r}$ = 0.001 (b) $p_{r}$ = 0.01 (c) $p_{r}$ = 0.1 (d) $p_{r}$ = 1}
		\label{R_w}
	\end{figure}

\begin{figure}[t]
		\centering
		\includegraphics[width= 0.5\textwidth]{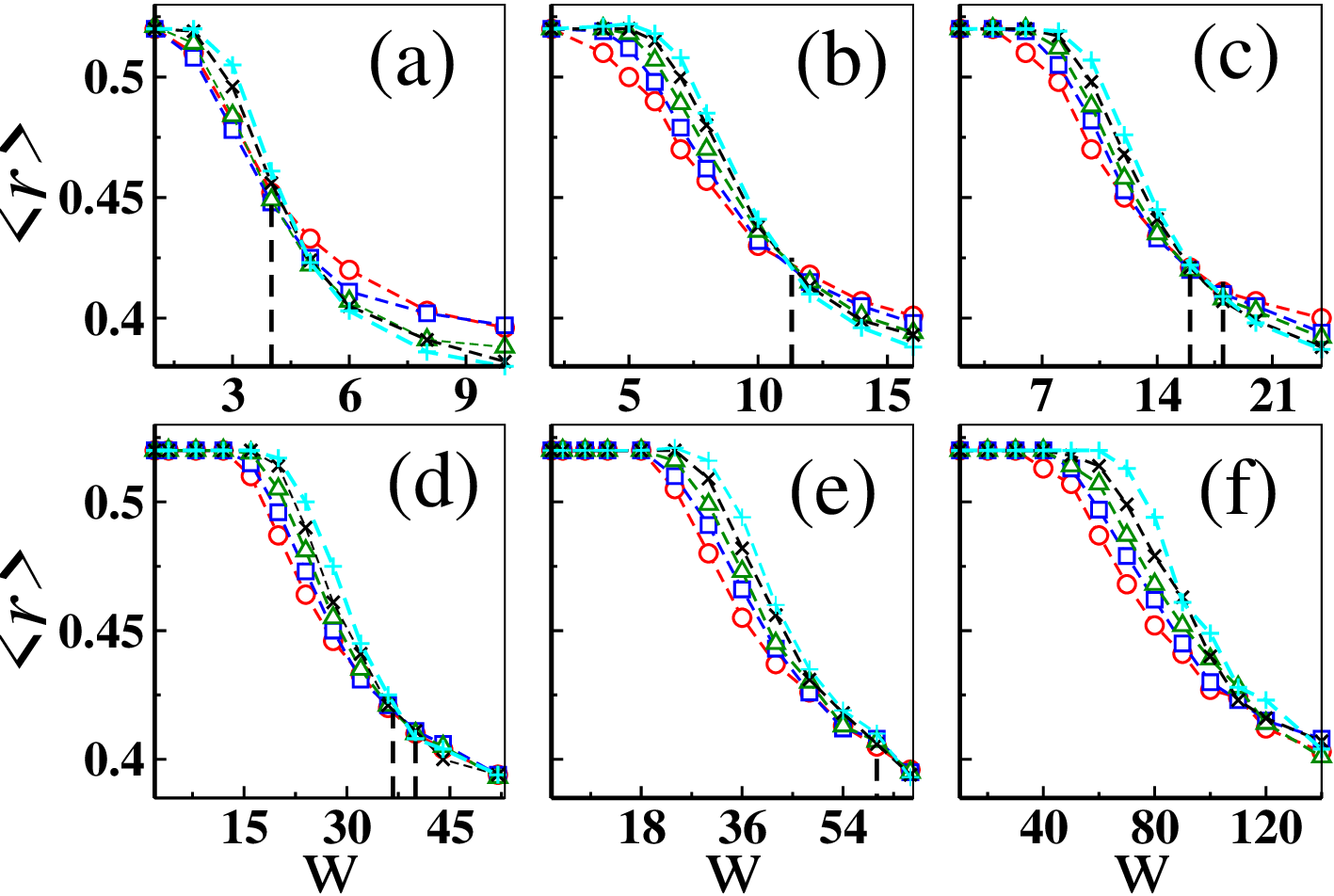}
		\caption{(Color online) Plot of $\langle r \rangle$ as a function of diagonal disorder ($w$) for various rewiring probabilities. $\color{red} \circ$, $\color{blue} \square$, $\color{green} \triangle$, $\color{black} *$ and $\color{cyan} +$ symbols are used for $N = 2000, 4000, 8000 $, $16000$ and $32000$ respectively with $\langle k \rangle = 20$. (a) $p_{r} = 0.001$ (b) $p_{r} = 0.005$ (c) $p_{r} = 0.01$ (d) $p_{r} = 0.05$ (e) $p_{r} = 0.1$ (f) $p_{r} = 1$.
			Vertical lines represents crossing point of $\langle r \rangle$ for different $N$ }
		\label{ravg_w}
	\end{figure}

	\paragraph{Disorder vs. randomness:}
	Let us now discuss results when the diagonal disorder is introduced in the adjacency matrix. 
	An increase in  $w$ leads to a gradual transition from the GOE to the Poisson statistics, as depicted in  Fig.~\ref{R_w}. However, the critical disorder $w_{c}$ required to achieve this transition increases with the increase in the value of $p_r$. The change in $\langle r \rangle$ with a change in $w$ for various values of $p_r$ and $N$ is plotted in Fig.~\ref{ravg_w}. $\langle r \rangle$ changes its value from $\langle r \rangle$ $\approx$ $0.52$ for $w = 0$ to  $\langle r \rangle$ $\approx$ $0.38$ for $w$ $=$ $w_{c}$. However, the value of $w_{c}$ increases with an increase in $p_{r}$. 
	In fact, for higher rewiring probabilities, $p_{r} (\geq 0.1$), it requires a much higher value of $w_{c}$.
	To obtain exact $w_{c}$, finite-size scaling analysis (FSS) would be required since critical phenomenon is defined only in the thermodynamic limit ($N$ $\rightarrow \infty$). The crossing point of the order parameter should remain the same with a change in the system size \cite{RMT_locn,T_2}. However, as argued in \cite{N_2}, for $d \rightarrow \infty$ ($l \sim ln(N)$), finite-size scaling analysis is nontrivial for many systems, for example, random regular or tree-like graphs, and one does not witness any crossing point. The order parameter $\langle r \rangle$ for such cases keeps drifting towards the Poisson statistics with increasing $N$. Since for small-world networks, $d \rightarrow \infty$, we are not performing any FSS analysis as it would required networks with very large sizes as done in \cite{SW_diff_model}. Nevertheless, we present the results for  $N \leq 32000$, demonstrating a similar trend  as in Fig.~\ref{ravg_w}.
	
	\begin{figure}[t]
		\centering
		\includegraphics[width= 0.5\textwidth]{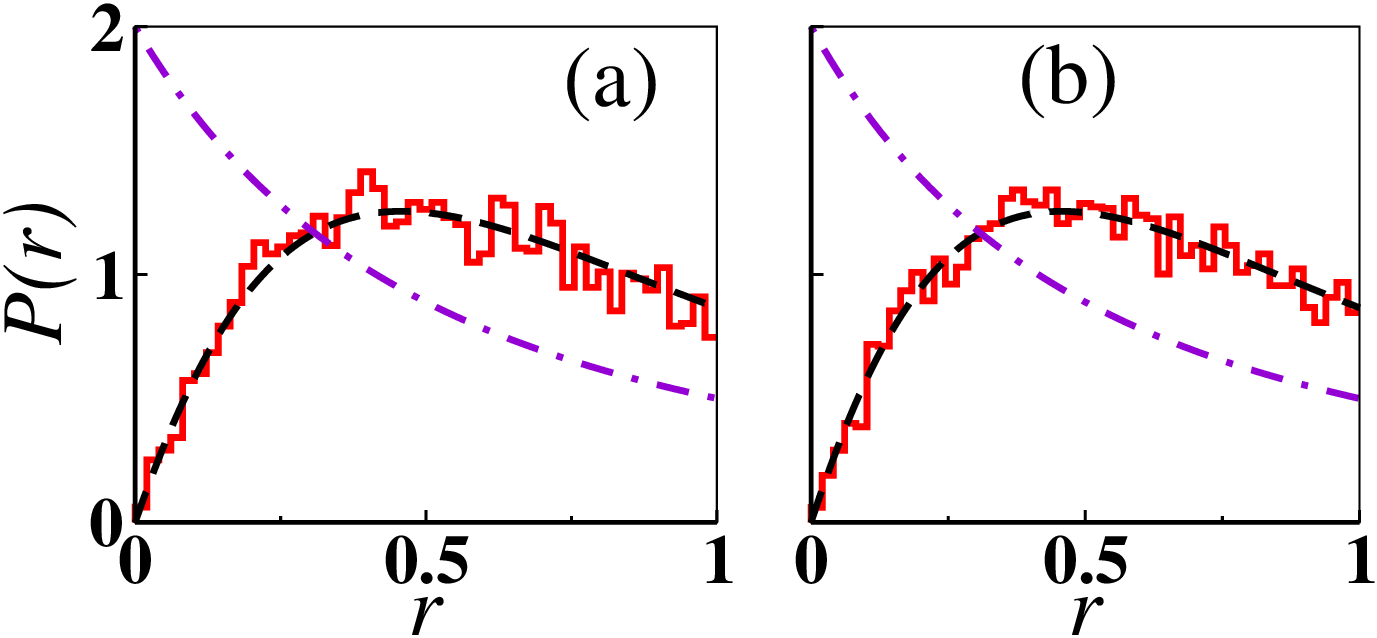}
		\caption{(Color online) Distribution of ratio of consecutive eigenvalue spacing with $w = 20$ from uniform distribution of width $2w$ for various probabilities of adding $1s$ in off diagonal entries of the adjacency matrix of 1D lattice. Black (dashed) and violet (dashed-dotted) lines, respectively indicate distribution function from Eq.~\eqref{eq:pr_goe} and \eqref{eq:poi} respectively. Here, $N = 1000$ and  $\langle k \rangle = 20$ with $20$ network realizations and $40$ disorder realizations. (a) $p = 0.1$ (b) $p = 0.2$ }
		\label{add_1}
	\end{figure}

	Further, rewiring affects the adjacency matrix of the initial 1D lattice ($p_{r}=0$) in two ways. First, there is a distortion in the diagonal band, and second is the random addition of $ones$ in the off-diagonal entries. Thus, we perform the following experiments to get an insight into which one of these two play a role in changing the eigenvalue ratio statistics. First, we keep the diagonal band undisturbed and randomly add $ones$ in the off-diagonal entries with a probability $p$. In this case,  the eigenvalue ratio statistics keeps depicting the GOE statistics irrespective of the probability  $p$ and the strength of the diagonal disorder (Fig.~\ref{add_1}). Here, results are shown only for one value of the diagonal disorder, but the same results for higher diagonal disorder strength have been obtained. 
	
\begin{figure}[t]
		\centering
		\includegraphics[width= 0.5 \textwidth]{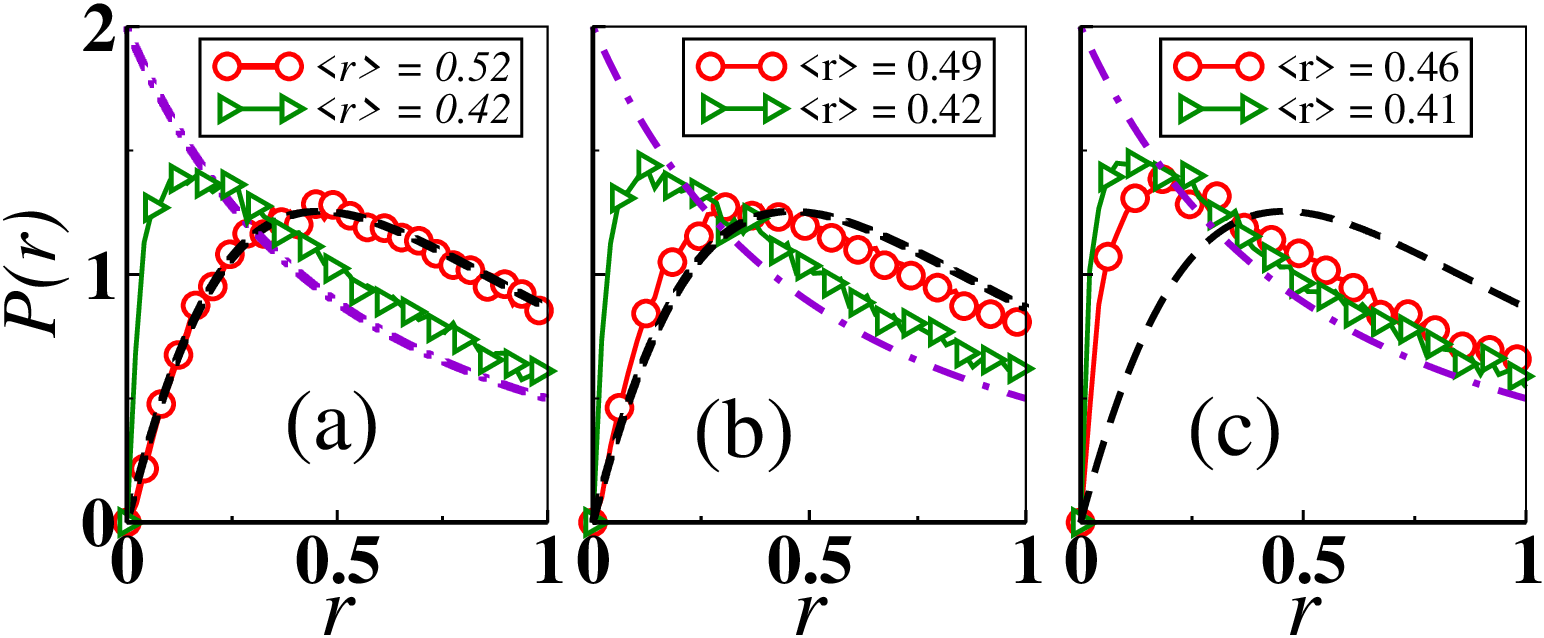}
		\caption{(Color online) Distribution of ratio of consecutive eigenvalue spacing with diagonal disorder from uniform distribution of width $2w$ for various probabilities of deleting $1s$ from diagonal band of the adjacency matrix of 1D lattice. $\color{red} \circ$ and $\color{green} \triangleright$ are used for $w = 1$ and $w = 4$ respectively. Black (dashed) and violet (dashed-dotted) lines, respectively indicate distribution function from Eq.~\eqref{eq:pr_goe} and \eqref{eq:poi} respectively. Here, $N = 1000$ and  $\langle k \rangle = 20$ with $20$ network realizations and $40$ disorder realizations. (a) $p = 0.001$ (b) $p = 0.01$ (c) $p = 0.1$}
		\label{omit_1}
	\end{figure}

	In the second experiment, we omit the $ones$ from the diagonal band uniformly with the probability $p$ and investigate its impact on the eigenvalue ratio statistics. Distorting the diagonal band with a  small probability ($p=0.001$) leads to a change from the GOE statistics for a  small disorder strength.\ In fact, for $p$ = $0.1$, even a very small value of $w$ leads to the Poisson statistics (Fig.~\ref{omit_1}).\ The above observation could be useful to explain the reason behind the increase in the critical value of disorder ($w_{c}$) with the increase in $p_r$. Small $p_r$ values yield small distortions in the diagonal band of the adjacency matrix accompanied with few filling up of $ones$ in the off-diagonal entries.\ This setup leads to the Poisson statistics for a small $w_{c}$ as omitting $ones$ even with a  small probability leads to the Poisson statistics. When rewiring probability is increased, though there is a distortion of the diagonal band, sufficient $ones$ have also been randomly distributed in the off-diagonal entries driving to the GOE statistics and thus higher $w_{c}$ is required. \\
	
	\paragraph{Impact of change in $d\lambda$:}
	
\begin{figure}[t]
		\centering
		\includegraphics[width= 0.50 \textwidth]{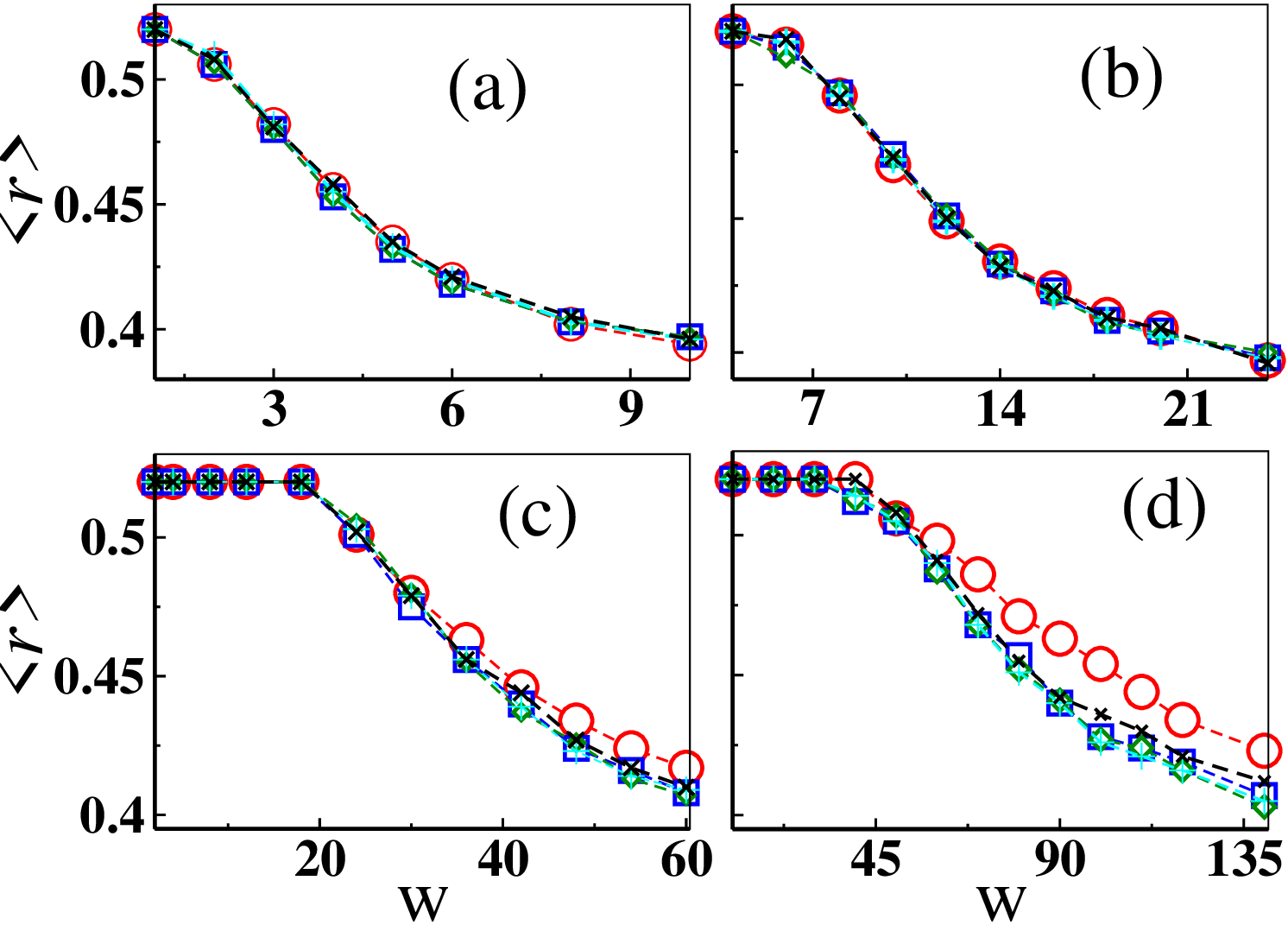} 
		\caption{(Color online) Plot of $\langle r \rangle$ as a function of diagonal disorder (w) and also with $d\lambda$ for various rewiring probabilities. $\color{red} \circ$, $\color{black} *$, $\color{blue} \square$, $\color {green} \Diamond$ and $\color{cyan} +$ are used for $d\lambda$ $\approx$ $0.10, 0.25, 0.50, 1.50$ and $3$ respectively. Here, $N = 2000$ and $\langle k \rangle = 20$. (a) $p_{r} = 0.001$ (b) $p_{r} = 0.01$ (c) $p_{r} = 0.1$ (d) $p_{r} = 1$}
		\label{ratio_delta}
	\end{figure}

	 Since we consider the eigenvalues in the central part of the eigenvalue spectrum, specifically those lying in the $d\lambda$ width around the $zero$ eigenvalue, let us discuss the rationale behind taking such a approach and an impact of $d\lambda$ on the results. First, the middle part of the spectrum is appreciably occupied, and the spacings become similar with different network sizes and thus help to reduce the finite-size effect to some extent \cite{dlam_1}. On the other hand, at the edges of a spectrum, the eigenvalues are not smooth for smaller $N$, but for $N \rightarrow \infty$, the eigenvalues spectrum becomes continuous, and thus the finite-size effect is more prominent. Second, to choose appropriate $d\lambda$, one has to pick an interval which is statistically sound and at the same time does not mix localized or delocalized eigenvectors for different values of $w$ \cite{dlam_2}. Fig.~\ref{ratio_delta} reflects that for $0.25 \leq d\lambda \leq 3$, a slight increase or decrease in the width does not have any noticeable impact on the value of $\langle r \rangle$ (for a given $w$) for all $p_r$ values. However, for $d\lambda \approx 0.10$, changes in $\langle r \rangle$ with respect to other $d\lambda$ values can be witnessed for higher rewiring probabilities ($p_{r} \geq 0.1$) (Fig.~\ref{ratio_delta}[c][d]). It is also important to note that as $w$ increases, there is a shortfall in the number of eigenvalues that are close to $zero$. In fact, we find that for the network parameters considered here, for $w > 70$, even $10^{4}$ random realizations yield a  total number of eigenvalues in the order of $10^{3}$ for $d\lambda \approx 0.10$. Thus, it is convenient to take $0.25 \leq$  $d\lambda$ $\leq 3$, in which $\langle r \rangle$ remains statistically sound for the same number of realizations for different $d\lambda$ values. Additionally, we checked using other measures like inverse participation ratio (not shown) that localized-delocalized eigenvectors do not mix in this range. We further add that a decrease in the value of $\langle k \rangle$ for a given $N$ may yield a quantitatively different impact for $d\lambda$, and there could be a larger number of eigenvalues in a given width $d\lambda$. \\

	\begin{figure}[t]
		\centering
		\includegraphics[width= 0.50 \textwidth]{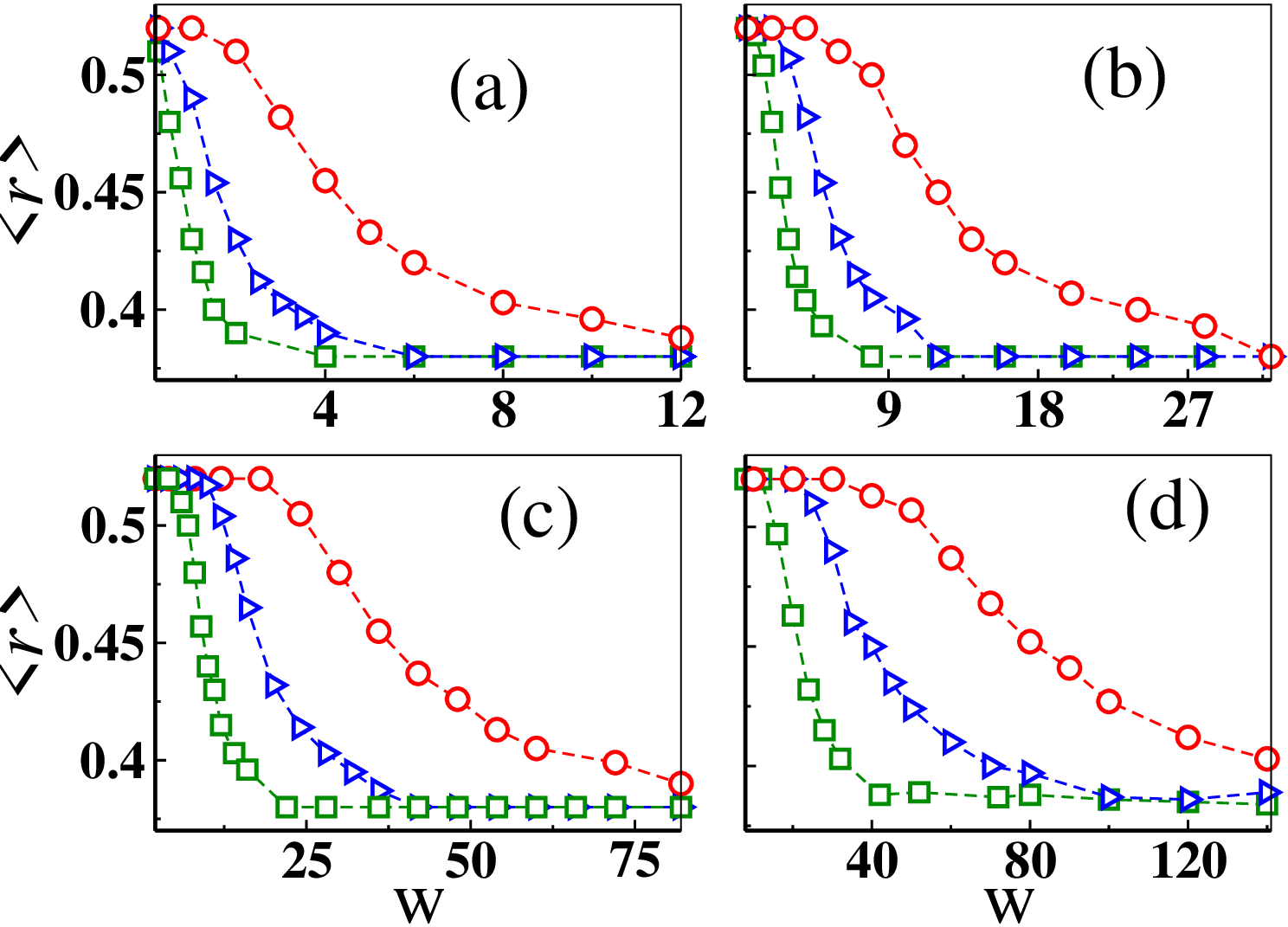} 
		\caption{(Color online) Plot of $\langle r \rangle$ as a function of diagonal disorder (w) and also with average degree $\langle k \rangle$ for various rewiring probabilities. $\color{red} \circ$, $\color{blue} \triangleright$ and $\color{green} \square$ are used for $\langle k \rangle$ $=$ $20,10$ and $6$ respectively. Here, $N = 2000$. 
			(a) $p_{r} = 0.001$ (b) $p_{r} = 0.01$ (c) $p_{r} = 0.1$ (d) $p_{r} = 1$}
		\label{ratio_sw_k}
	\end{figure}

	\paragraph {Impact of average degree:}
	We further probe the impact of average degree on the statistics of ratios of consecutive eigenvalue spacings. Note that the largest eigenvalue of the network is bound with the largest degree $k^{max}$ \cite{bulk_k}.  Moreover, for a random network $\lambda_{1}$ $\approx$ $\langle k \rangle$. Thus, varying $\langle k \rangle$ may affect the eigenvalue spectrum drastically even for fixed network size. In our analyses, we have considered, $\langle k \rangle=  6, 10,$ and $20$ with $N = 2000$ being fixed.
	It is also worth noting that a  decrease in the average degree may affect the probability ($p_{c}$) at which small-world transition occurs.
	However, we notice that reducing $\langle k \rangle$ from $20$ to $10$ does not affect $p_{c}$ and it remains equal to $0.01$.
	Fig.~\ref{ratio_sw_k} illustrates the change in $\langle r \rangle$ when  $\langle k \rangle$ is varied. It is apparent from the figure that when $\langle k \rangle$ is decreased, the critical disorder ($w_{c}$) required to procure the transition also decreases for all the values of the rewiring probabilities.

	\begin{figure}[t]
		\centering
		\includegraphics[width= 0.4\textwidth]{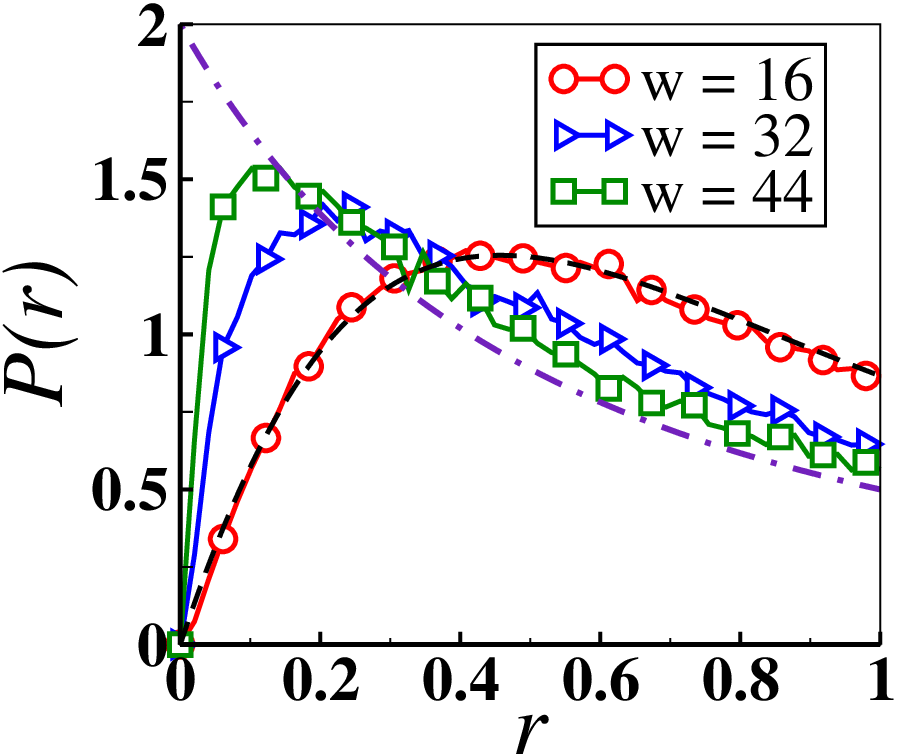} 
		\caption{(Color online) Distribution of ratio of consecutive eigenvalue spacing for ER random network. Black (dashed) and violet (dashed-dotted) lines, respectively indicate GOE and Poisson distribution from Eq.~\eqref{eq:pr_goe} and \eqref{eq:poi} respectively. $N = 2000$ and $\langle k \rangle = 8$ with $20$ network and $40$ disorder realizations.}
		\label{ratio_ER}
	\end{figure}

	\subsection{\bf Eigenvalue ratio statistics of ER random networks} \label{a4}
	We now extend the investigation to ER random networks. ER random networks are constructed using ER model \cite{ER} as follows. Starting with $N$ nodes and average degree $\langle k \rangle$, each pair of the nodes is connected with a probability $p=\langle k \rangle/N$. The degree distribution of ER random networks follows a binomial distribution. It is also worth noting that small-world random networks with $p_{r}$ $=$ $1$ and ER random networks are slightly different in the sense that while the former has fixed links with different realizations, it may change in the latter. 
	Additionally, the small-world random network ($p_{r} = 1$)
	retains the initial regular structure ($p_{r} = 0$) as its memory and degree distribution follow a normal distribution with peak at $\langle k \rangle$ and a small variance. On the contrary, in ER random network, though the degree distribution peak stays around $\langle k \rangle$, the variance is larger than the small-world random networks. It is evident from Fig.~\ref{ratio_ER}, similar to the small-world networks, the eigenvalue ratio statistics of ER random networks depict GOE to the Poisson transition with an increase in the diagonal disorder.

	\paragraph {Assortative-disassortative networks:} 
	The degree-degree correlation is one of the key characteristics of real-world networks \cite{ass_1}. In many real-world networks, like social networks, a node with a high degree tends to connect with similar high degree nodes, commonly known as assortative networks \cite{ass_4}. This characteristic of networks is known as assortativity or assortative mixing. On the other hand, in biological and technological networks, high degree nodes prefer to connect with the nodes having a low degree, referred to as disassortative networks, and the property is known as disassortativity, or disassortative mixing \cite{ass_2}. To incorporate assortative or disassortative mixing in the original ER random network, we use the reshuffling algorithm \cite{re_shufl}. The degree of  (dis)assortativity is quantified by the Pearson (degree-degree) correlation coefficient, denoted as $r_{a}$ where $-1$ $\leq$ $r_{a}$ $\leq$ $1$.
	For most assortative network $r_{a}$ will be closer to 1 while for most disassortative network, $r_{a}$ will be close to  $-1$. Note that the probability distribution of eigenvalues (spectral density) changes drastically with the change in $r_{a}$ \cite{alok}. Thus, it would be interesting to probe eigenvalue ratio statistics for the  (dis)assortative ER network.

	\begin{figure}[t]
		\centering
		\includegraphics[width= 0.5\textwidth]{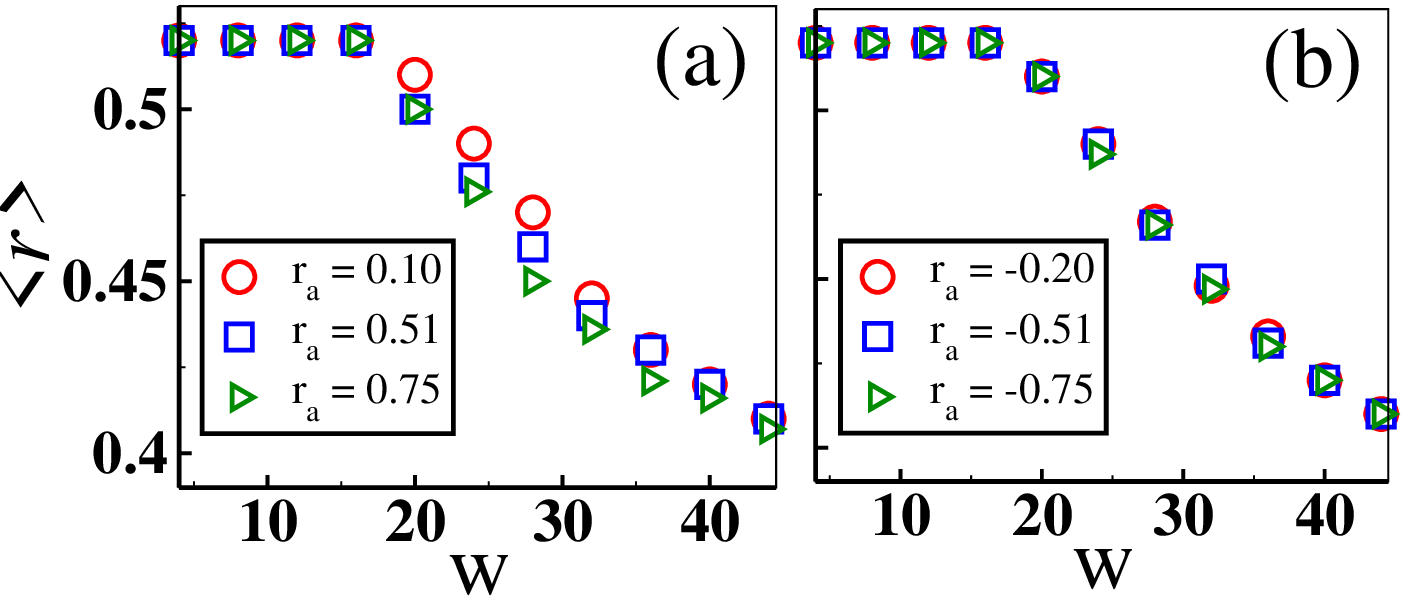} 
		\caption{(Color online) $\langle r \rangle$ is plotted as a function of $w$. (a) Assortative ER networks (b) Disassortative ER networks. Here, we have considered $N = 2000$ and  $\langle k \rangle = 8$ with $20$ network and $40$ disorder realizations.  }
		\label{avg_r_er_ass}
	\end{figure}

	We considered $N = 2000$ and $\langle k \rangle = 8$ in our analyses. The average degree is kept at this small value to ensure GOE to Poisson transition for a finite value of $w$ otherwise, for larger  $\langle k \rangle$, $w$ will be far-reaching.\ First, we study the distribution of the ratio of consecutive eigenvalue spacing without the diagonal disorder in the adjacency matrix.
	We find that it follows GOE statistics irrespective of the value of $r_{a}$.
	We next introduce diagonal disorder in the adjacency matrix and study eigenvalue ratio statistics. We find that
	if the degree of assortativity ($0 \leq r_{a} \leq 1$) is increased, though there is no significant change in $w_{c}$, $\langle r \rangle$ shows slightly lesser values as compared with the corresponding less assortative networks (Fig.~\ref{avg_r_er_ass} [a]). On the other hand, with an increasing degree of disassortativity ($-1 \leq r_{a} < 0$), we do not find its effects on eigenvalue ratio statistics (Fig.~\ref{avg_r_er_ass} [b]). It is important to note here that upon increasing the degree of assortativity leads to a decrease in the randomness, as discussed in \cite{alok}.
	Moreover, changing the degree of disassortativity does not affect the randomness
	in networks \cite{alok}, which is also reflected in our analysis as for any $w$, $\langle r \rangle$ remains the same with a change in $r_{a}$.
	We want to stress here that randomness induced in small-world networks upon links rewiring is more notable as compared with that brought upon by the (dis)assortative mixing in the ER network \cite{SJ_epl,alok}. Hence, variation in $w_{c}$ as a function of $p_{r}$ is more significant.

	\subsection{\bf Maximal entropy random walk (MERW)} \label{a5}
	Localization of eigenvectors of the  adjacency and Laplacian matrices is known to influence various dynamical processes on the corresponding networks. For example, localization of the principal eigenvector  of a adjacency matrix is known to play a pivotal role in disease spreading \cite{PEV_disease}, perturbation propagation in ecological networks \cite{PEV_ecological}, etc. However, the exact underlying mechanism remains elusive, particularly such understandings for the non-principal eigenvalues and eigenvectors are missing. Though  most of the spectral investigations have revolved around the principal eigenvectors and the corresponding eigenvalue, sporadic investigations indicate that non-principal eigenvectors and associated eigenvalues of the adjacency matrices of networks contribute to the transient relaxation dynamics \cite{TD_1, TD_2}.

	This section studies dynamics of maximal entropy random walker (MERW) in various model networks. It then relates its dynamics with the localization (Poisson statistics) and delocalization (GOE statistics) properties of the underlying model networks. MERW was first introduced in \cite{merw_1} where it was argued that MERW localizes in few nodes, which is not with the case of generic random walk (GRW). To begin with, we first discuss a general framework for the maximal entropy random walk.
	Let us consider a random walker hopping from node to node on a connected, undirected, unweighted graph {\it G} $=$ \{V,E\}.\ 
	At each time step, a walker sitting at any node, say $i$, jump to its neighboring node $j$ with a probability $P_{ij}$ indicating the probability of jumping of the random walker from the node $i$ to the node $j$ independent of the previous history. Note that, $P_{ij} = 0$, if $A_{ij} = 0$, since a walker can jump to its neighbouring nodes only. The elements of the transition matrix $P$ can be determined as,
	\begin{equation}
	P_{ij} = \frac{A_{ij}}{\lambda_{1}}\frac{\psi_{j}} {\psi_{i}}   
	\end{equation}

	where $\lambda_{1}$ is the largest eigenvalue of the corresponding adjacency matrix and $\psi_{j}$ and $\psi_{i}$, respectively, are the $j^{th}$ and $i^{th}$ components of the normalized principal eigenvector. The Perron-Frobenius theorem states that all the elements of the principal eigenvector have the same sign, so that $P_{ij} \geq 0$.
	Also, one can easily see that for each node $i$, $\sum_{j} P_{ij} = 1$.
	One quantity of interest is probability of finding the walker at any node $i$ at a  time $t$ denoted as $p_{i}(t)$. One can easily compute $p_{i}(t+1) = \sum_{j} p_{j}(t)  P_{ji} $. After a time $t$, $p_{i}(t)$ will reach to a steady state, $p_{i}^{*} = \sum_{j} p_{j}^{*}  P_{ji} $. One of the most important properties of MERW is that for a given length $t$ and a pair of the end points, say, walker started from node $i_{o}$ and ends at $i_{t}$, all trajectories are equiprobable which is not  the case of generic random walkers. For a generic random walk (GRW), $P_{ij}$ $=$ $\frac{A_{ij}}{k_{i}}$ where $k_{i}$ is the degree of the $i^{th}$ node. Note that in GRW, trajectories for given length $t$ and given endpoints $i_{o}$, $i_{t}$, are not equiprobable. Let $\vec p (t)$ $= \{p_{1}(t), p_{2}(t), p_{3}(t), $\ldots$, p_{N}(t)\}$ be the probability distribution of a random walker on a given network. For a given initial probability distribution of the walker on the graph, ($\vec p (0)$) and transition matrix $P$,  one can easily compute the probability distribution of the walker at any time $t$ as,
	\begin{equation}
	\vec p (t) =\vec p (0) P^{t}
	\end{equation}
	Next, we consider the Shannon entropy $S$ of the walker at each time step $t$, as 
	\begin{equation}
	S(t) = -\sum_{i} p_{i}(t) ln (p_{i}(t))
	\end{equation}
	Note that, $S \rightarrow 0$, if the walker is sitting at one node only, say, $i_{o}$; $p_{i_{o}} = 1$ and $p_{i} = 0$ for all the  other nodes.  On the other hand, if the probability of finding the walker is equal at each node, $p_{i}$ $=$ $1/N$, $S \rightarrow ln(N)$. Thus, $0 \leq S \leq ln(N)$. Also, in large time $t$,  $S(t)$ will reach to the steady state, and $S(t+1)$ $=$ $S(t)$ $\approx$ $c$, where $c$ is some constant. In the steady state, probability of finding the walker at different nodes does not change with time. We denote $\tau$ be the time taken by the walker to reach the steady state starting from a given initial condition.
	We evolve the random walker following MERW transition matrix on the following model networks; small-world, ER random, and (dis)assortative ER networks. We choose two initial conditions to avoid any predilection in our analysis.  (1) In the first case, we choose homogeneous probability distribution for the walker, i.e., the probability of finding the walker at each node is equal to $1/N$ at $t = 0$.  We then evolve the walker for a sufficiently long time until it reaches the steady state and calculate $S$ for each time step. (2) We randomly choose a node, say $i_{0}$, which acts as the starting point for the random walker with a probability $1$.
	
	\begin{figure}[t]
		\centering
		\includegraphics[width= 0.5\textwidth]{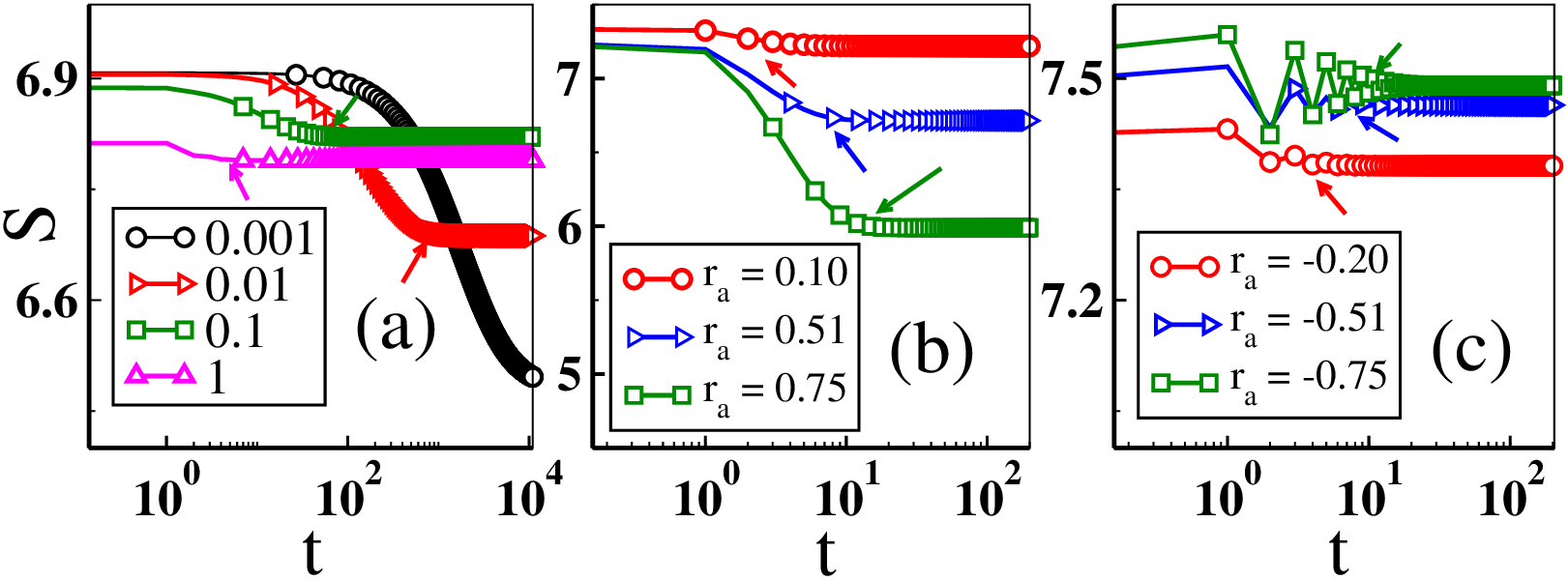} 
		\caption{(Color online) $S$ as a function of $t$ for various model networks when initial condition is  homogeneous distribution, (a) Small-world networks, (b) assortative ER networks, and (c) disassortative ER networks. Here, (a) $N = 1000$ and  $\langle k \rangle = 10$ and (b)-(c) $N = 2000$ and $\langle k \rangle = 8$. Arrow indicate position where $S$ hits the steady state.}
		\label{merw_homg}
	\end{figure}
	
	\begin{figure}[t]
		\centering
		\includegraphics[width= 0.5\textwidth]{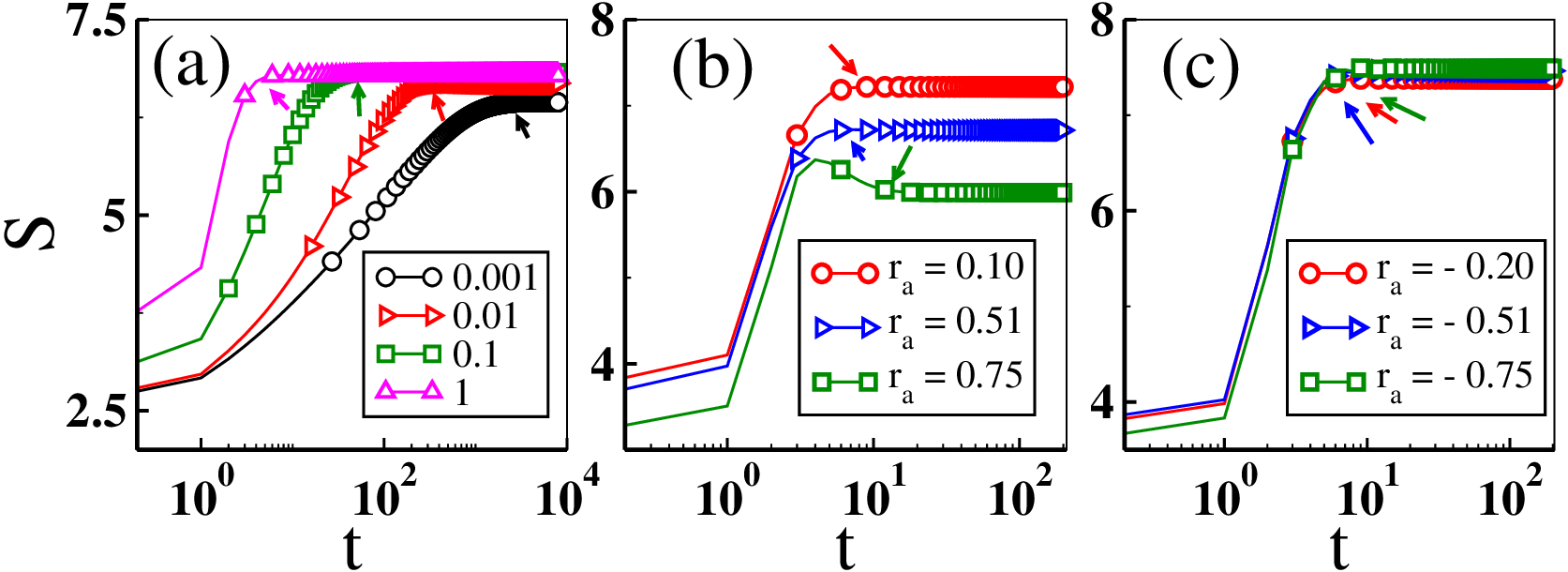} 
		\caption{(Color online) $S$  as a function of $t$ for various model networks when random walker starts from a randomly chosen node with probability $1$; (a) small-world networks, (b) assortative ER networks, and (c) disassortative ER networks. Other parameters are same as Fig.~10. }
		\label{merw_locn}
	\end{figure}

	We first discuss the results for small-world networks. As discussed in the previous section that the critical disorder ($w_{c}$) required to procure the Poisson statistics shows an increase with an increase in the value of $p_r$. Figs.~\ref{merw_homg}(a), and \ref{merw_locn}(a) represent the entropy of the random walker against time $t$ for different values of $p_r$ of small-world networks. The time $\tau$ after which the random walker reaches the steady state, i.e., $S \rightarrow c$  for $t > \tau$ decreases with the increase in $p_r$ for both cases of the initial conditions. Thus, $\tau \propto 1/w_{c}$.
	This is a crucial observation that may provide important insight into the localization of the random walker. When the walker starts from a randomly chosen node, say $i_{o}$, for smaller $t$, the probability of finding the walker is finite only for the nodes which are neighbors of $i_{0}$. However, for larger $t$, there is a finite probability of finding the walker on any node. Further, for $t \geq \tau $, $p_{i} > 0$ $\forall i \in\left\{1,2,3 \ldots ,N\right\}$ and does not change with time. Next, for smaller values of the rewiring probability, $\tau$ is very high, and thus even after a larger $t$, the probability of finding the walker on most of the nodes remains $p_{i} \rightarrow 0 $. On the contrary, for higher values of the rewiring probability, $\tau$ is very small. Hence, the probability of finding the walker on all the nodes of the network becomes finite even for a shorter $t$. Therefore, the probability of finding the walker being finite for all the nodes (for small $t$), we would need a higher disorder strength to localize it on a limited set of the nodes, as is the case for the high $p_r$ values. On the other hand,  if the probability of finding the walker remains nonzero only for the few nodes even after long time, a low diagonal disorder strength would be enough to localize the walker as it still remains in the purview of few nodes, which is the case of small $p_r$ values. The evolution of the probability distribution of walker on small-world networks can be found in the supplementary material \cite{SM}.

	We further extend this analyses to the (dis)assortative ER networks. 
	As apparent from Figs.~\ref{merw_homg} (b) and \ref{merw_locn} (b),  the value of $\tau$ increases with assortativity which is consistent with the earlier observation of $\tau \propto 1/w_{c}$.  Further, for disassortative networks,
	as discussed earlier, there exists no visible effect of disassortativity ($-1 \leq r_{a} < 0 $) on the eigenvalue ratio statistics (Fig.~\ref{avg_r_er_ass}[b]). From Figs.~\ref{merw_homg} [c],  \ref{merw_locn} [c], it is visible that the
	change in the value of $\tau$ remains insignificant as $\langle r \rangle$ is unchanged with the change in value of $r_{a}$ for disassortative networks.

	\section{Conclusion} \label{a6}
	To conclude, we studied the eigenvalue ratio statistics of various model networks. 
	For the small-world networks, we find that as the strength of diagonal disorder increases, the eigenvalue ratio distribution depicts a gradual transition from the GOE to the Poisson statistics. However, the critical disorder ($w_{c}$) required to obtain the transition increases with the increase in the value of the rewiring probability. Thus, higher $w_{c}$ is required to obtain GOE to Poisson transition when randomness in the network increases.
	Next, we analyzed the impact of change in the network's average degree on the eigenvalue ratio statistics.
	As expected, a decrease in the average degree leads to a decrease in the value of the critical disorder required to induce Poisson statistics.
	Next, we extend our analysis to the ER random networks. In this case, also, we found the gradual transition of GOE to Poisson statistics upon the introduction of diagonal disorder. Finally, to check the effect of degree-degree correlation, we perform (dis)assortative mixing in the original ER random network. Interestingly, we find that an increase in the degree of assortativity leads to a slight decrease in the value of $\langle r \rangle$ for a  given $w$. On the other hand, when degree of disassortativity increases, there is no noticeable  impact on the eigenvalue ratio statistics.
	Further, we relate the value of the critical disorder ($w_{c}$) with the time taken by the  maximal entropy random walker to reach to the steady state. The lower the  $w_{c}$ (for fixed $N$ and $\langle k \rangle$), the higher time is taken by the walker to reach to the steady state. Further, we argued that when the walker has reached to the steady state,
	the probability of finding it on all the nodes becomes finite. For small $\tau$, the walker would be able to access all the nodes in a sufficiently shorter time, and thus it requires a high value of the diagonal disorder strength to make the network localized. On the other hand, for sufficiently longer $\tau$,  probability of finding the walker remains finite on a few nodes, and consequently, a low $w_{c}$ would be enough to make it localized.
	
	 Few previous studies have investigated the implications of extremal eigenvectors localization of adjacency matrices of networks for various dynamical behaviour of corresponding system. For example, in \cite{PEV_disease}, it was shown that if  infection rate is slightly higher than the threshold and the principal eigenvector is localized, the disease will be localized on a finite set of vertices. Also, recently in \cite{Econ_1}, the importance of localization of eigenvector corresponding to $\lambda_{min}$ was discussed. The authors argued that the stability of the system will depend on the localization  nature of the eigenvector corresponding to $\lambda_{min}$. Furthermore, importance of the spread of the bulk part of the eigenvalues spectrum with an increase in the diagonal disorder for steering localization behaviour of eigenvector corresponding to $\lambda_{min}$ was also argued by the authors. Additionally, in \cite{APL_RMT_1, APL_RMT_2}, communities in real-world networks were characterized/identified using RMT and properties of highly localized eigenvectors.
		However, the exact applications of non-principal eigenvalues and corresponding eigenvectors is still missing in the network science literature except that they are  known to be contributing in transient dynamics.\ In this work, we analyze the localization-delocalization transition using the eigenvalue ratio statistics and its implications for the maximally entropy random walkers. The eigenvalue ratio statistics is  already a popular technique in condensed matter research for capturing the localization-delocalization transition in various systems, and in the regular graphs. We anticipate that the eigenvalue ratio statistics has the same scope in network science and can be used to capture or quantify different phase transitions, as well as to get insight to various dynamical processes like disease spreading, random walker, evolutionary dynamics etc. We further expect that our work can be applied on the systems having diagonal disorders, such as in \cite{Econ_1,Apl_2, Apl_3} which can be characterized by underlying network structure. Further, this paper only considers homogeneous  networks, and there  exists a vast number of real-world networks having heterogeneous degree distributions, for example scale-free. A straight forward and interesting future  direction is to extend the present framework for  scale-free networks having different degree mixing (assortative, disassortative) properties.

	\section {Acknowledgements}
	S.J. acknowledges Govt of India, BRNS  Grant  No.  37(3)/14/11/2018-BRNS/37131 for financial support. SJ acknowledges DST POWER Grant SERB/F/9035/2021-2022. We thank Santosh Kumar (Shiv Nadar University) for  useful suggestions.

\providecommand{\noopsort}[1]{}\providecommand{\singleletter}[1]{#1}%

\providecommand{\noopsort}[1]{}\providecommand{\singleletter}[1]{#1}%

\end{document}


\title{Supplementary Material:Eigenvalue ratio statistics  of complex networks: Disorder vs. Randomness}
\author{Ankit Mishra, Tanu Raghav and Sarika Jalan}
\affiliation{Complex Systems Lab, Department of Physics, Indian Institute of Technology Indore, Khandwa Road, Simrol, Indore-453552, India}

\maketitle

\subsection{Eigenvalue ratio statistics for higher diagonal disorder strength}

\begin{figure}[ht]
	 	\centering
	 	\includegraphics[width= 0.45\textwidth]{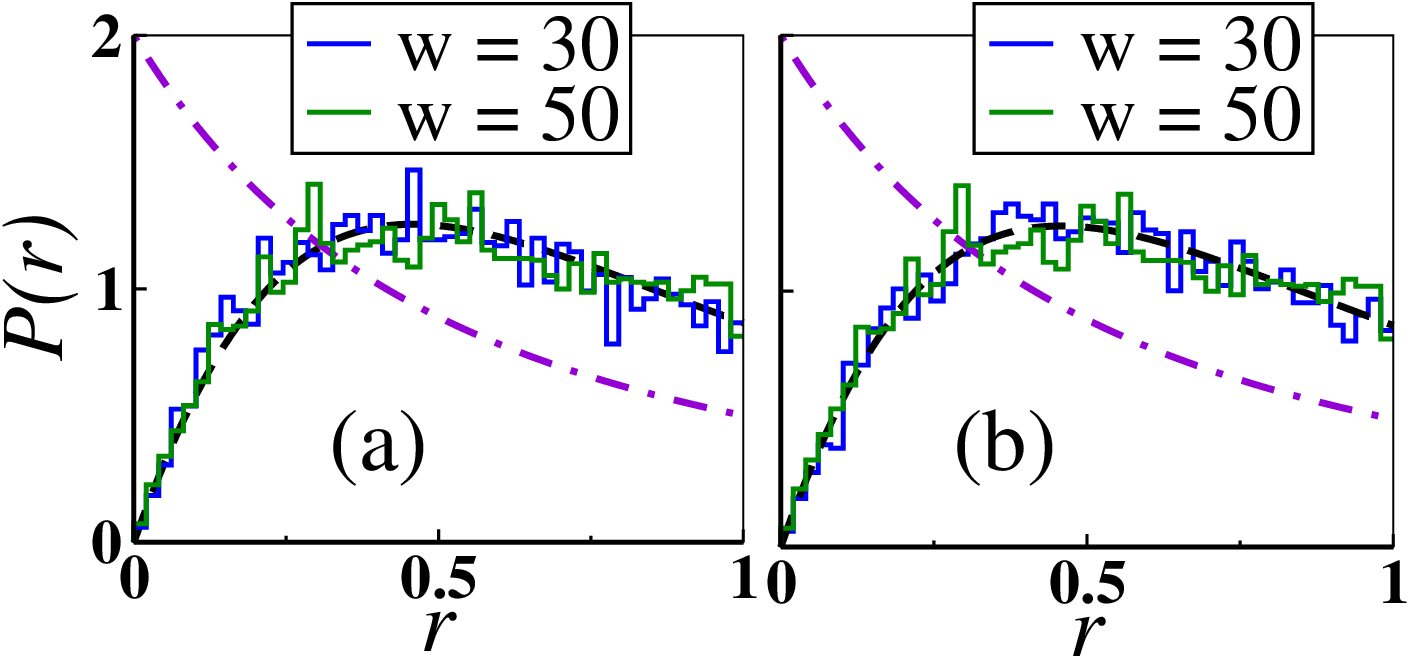}
	 	\caption{ Distribution of ratio of consecutive eigenvalue spacing with diagonal disorder from uniform distribution of width $2w$ for various probabilities of adding $1s$ in off diagonal entries of the adjacency matrix of 1D lattice. Black and Violet line indicate distribution function of GOE and Poisson statistics respectively.
		Here, $N = 1000$ and  $\langle k \rangle = 20$ with $20$ network realizations and $40$ disorder realizations. 
	(a) $p = 0.1$ (b) $p = 0.2$ }
	 	\label{add_1_HW}
	 \end{figure}

Fig.~\ref{add_1_HW} demonstrate the eigenvalue ratio statistics for the adjacency matrix when the diagonal band is preserved and $ones$ are randomly added in the off-diagonal with probability $p$. The eigenvalue ratio keep depicting GOE statistics even for higher diagonal disorder strength. 

\subsection{Maximal Entropy Random Walk (MERW)} 

\begin{figure}[ht]
	\centering
	\includegraphics[width= 0.55\textwidth]{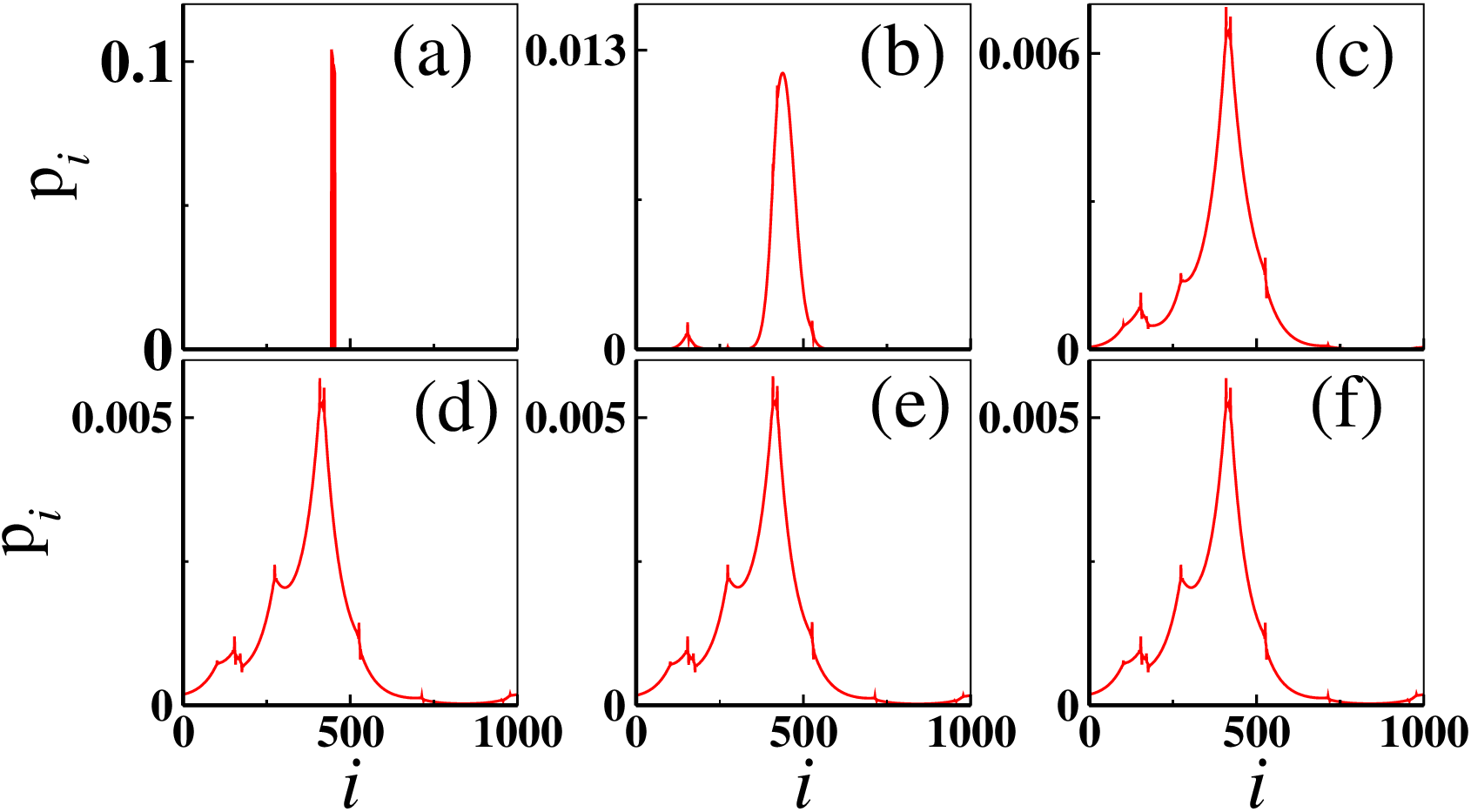} 
	\caption{ Probability of finding the walker on different nodes for small-world networks with $p_{r} = 0.001$. (a) $t = 1$, (b) $t = 100$, and (c) $t = 1000$, (d) $t = 2000$ (e) $t = 5000$ (f) $t = 10000$;
			$N = 1000$ and $\langle k \rangle = 8$ }
	\label{S1}
\end{figure}

Fig.~\ref{S1} demonstrate the evolution of probability distribution of finding the  walker on different nodes for small-world networks with $p_{r} = 0.001$. It is visible from the figure that even after long time $t$, the probability of finding the walker remains finite only for few nodes. Thus, a small disorder strength would be enough to make the walker localized on finite nodes. \\

\begin{figure}[ht]
	\centering
	\includegraphics[width= 0.55\textwidth]{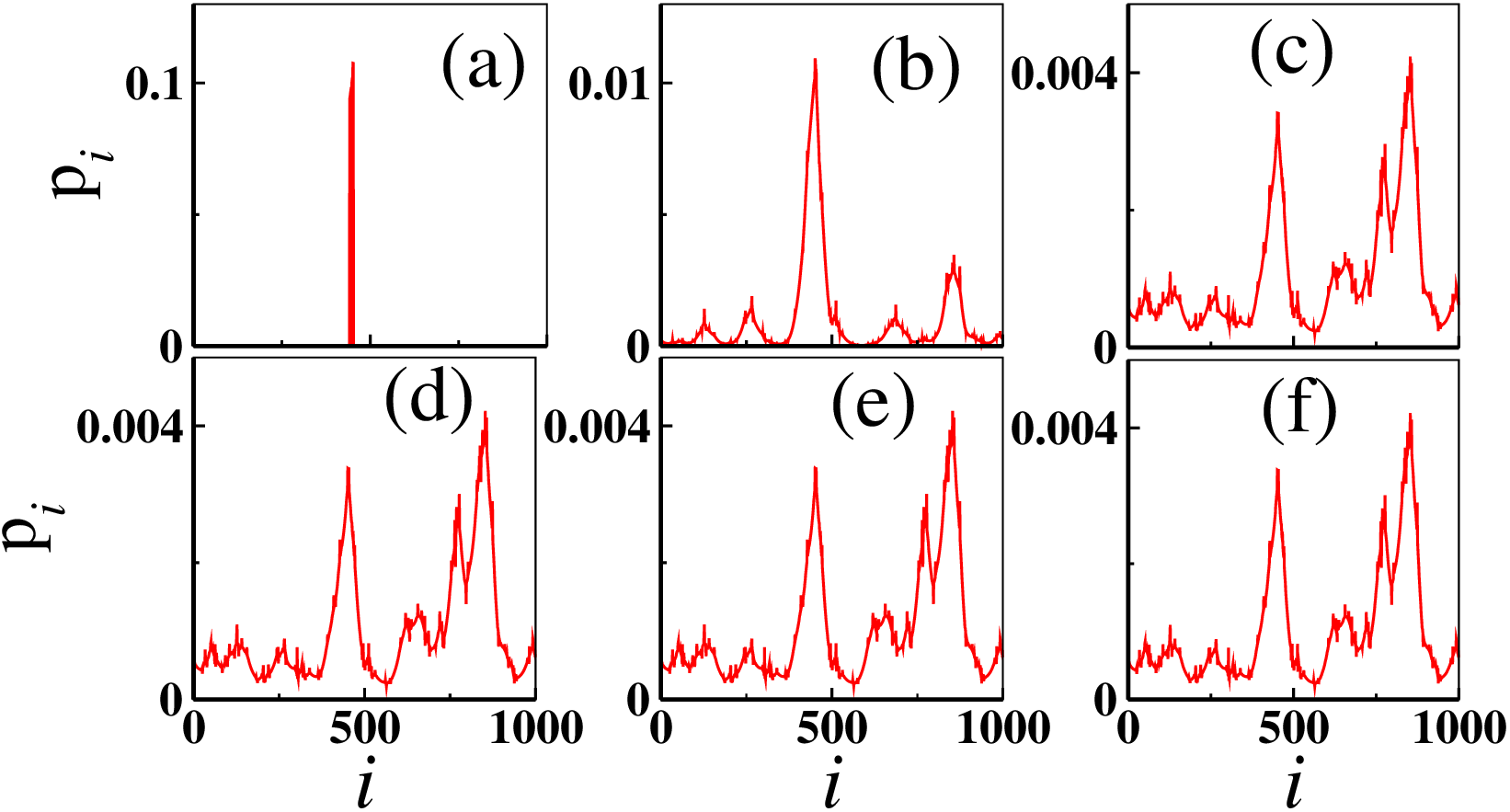} 
	\caption{ Probability of finding the walker on different nodes for small-world networks with $p_{r} = 0.01$. (a) $t = 1$, (b) $t = 100$, and (c) $t = 1000$, (d) $t = 2000$ (e) $t = 5000$ (f) $t = 10000$;
	$N = 1000$ and $\langle k \rangle = 8$ }
	\label{S2}
\end{figure}

Fig.~\ref{S2} demonstrate the evolution of probability distribution of finding the walker on different nodes for  small-world networks with $p_{r} = 0.01$. In this case, after $t >1000$ probability of finding the walker on all the nodes becomes finite. Since, $\tau$ is less compared with the $p_{r} = 0.001$, eventually a high disorder strength would be needed for localization.\\

\begin{figure}[ht]
	\centering
	\includegraphics[width= 0.55\textwidth]{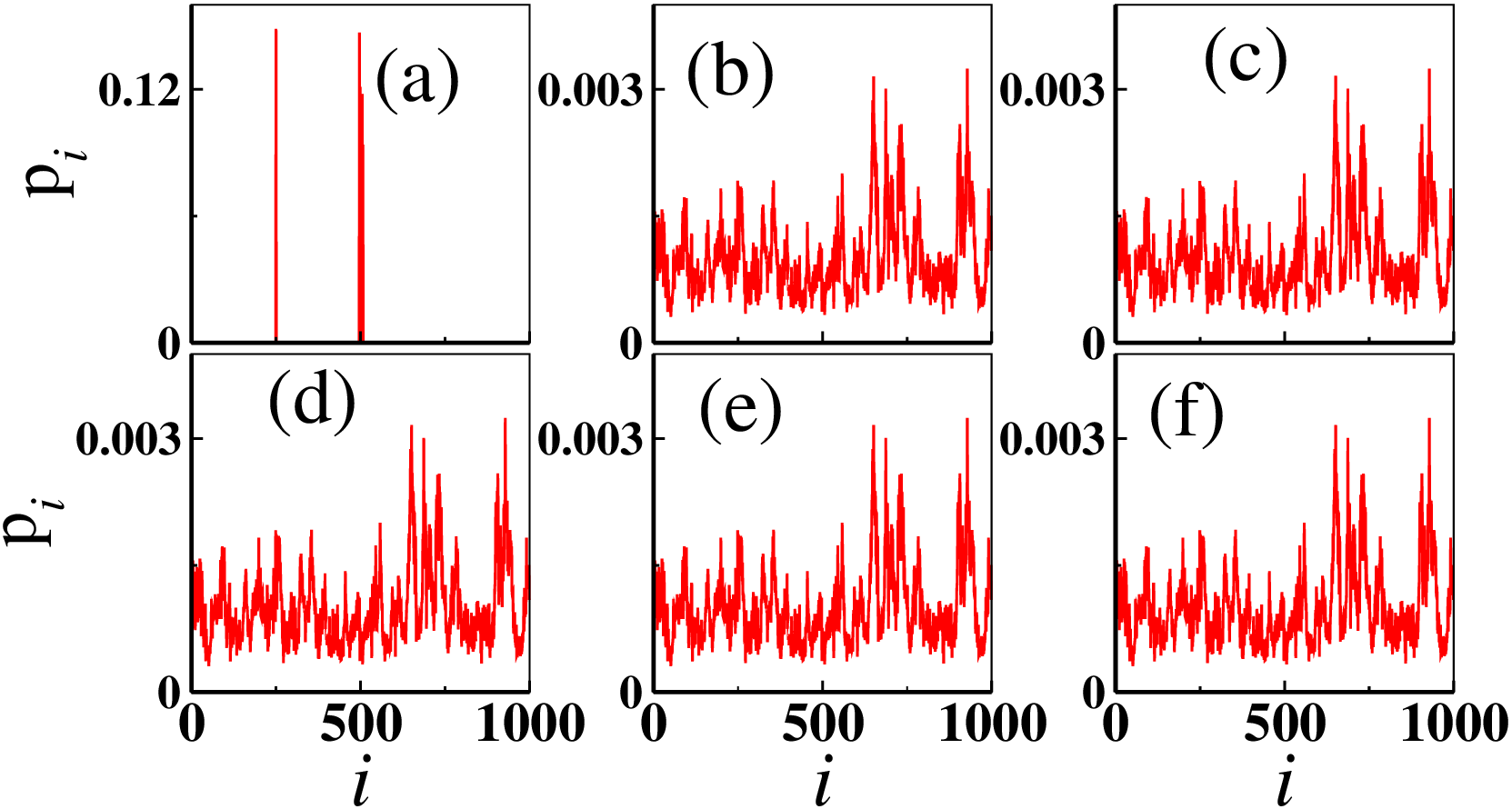} 
	\caption{ Probability of finding the walker on different nodes for small-world networks with $p_{r} = 0.1$. (a) $t = 1$, (b) $t = 100$, and (c) $t = 1000$, (d) $t = 2000$ (e) $t = 5000$ (f) $t = 10000$;
			$N = 1000$ and $\langle k \rangle = 8$}
	\label{S3}
\end{figure}

Fig.~\ref{S3} demonstrate the evolution of probability distribution of finding the  walker on different nodes for  small-world networks with $p_{r} = 0.1$. Here, the walker reached the steady state for a shorter $t$, and the probability of finding the walker becomes finite for all the nodes. Hence, to localize the walker, a high disorder strength would be required compared with the small rewiring probabilities.\\

\begin{figure}[ht]
	\centering
	\includegraphics[width= 0.55\textwidth]{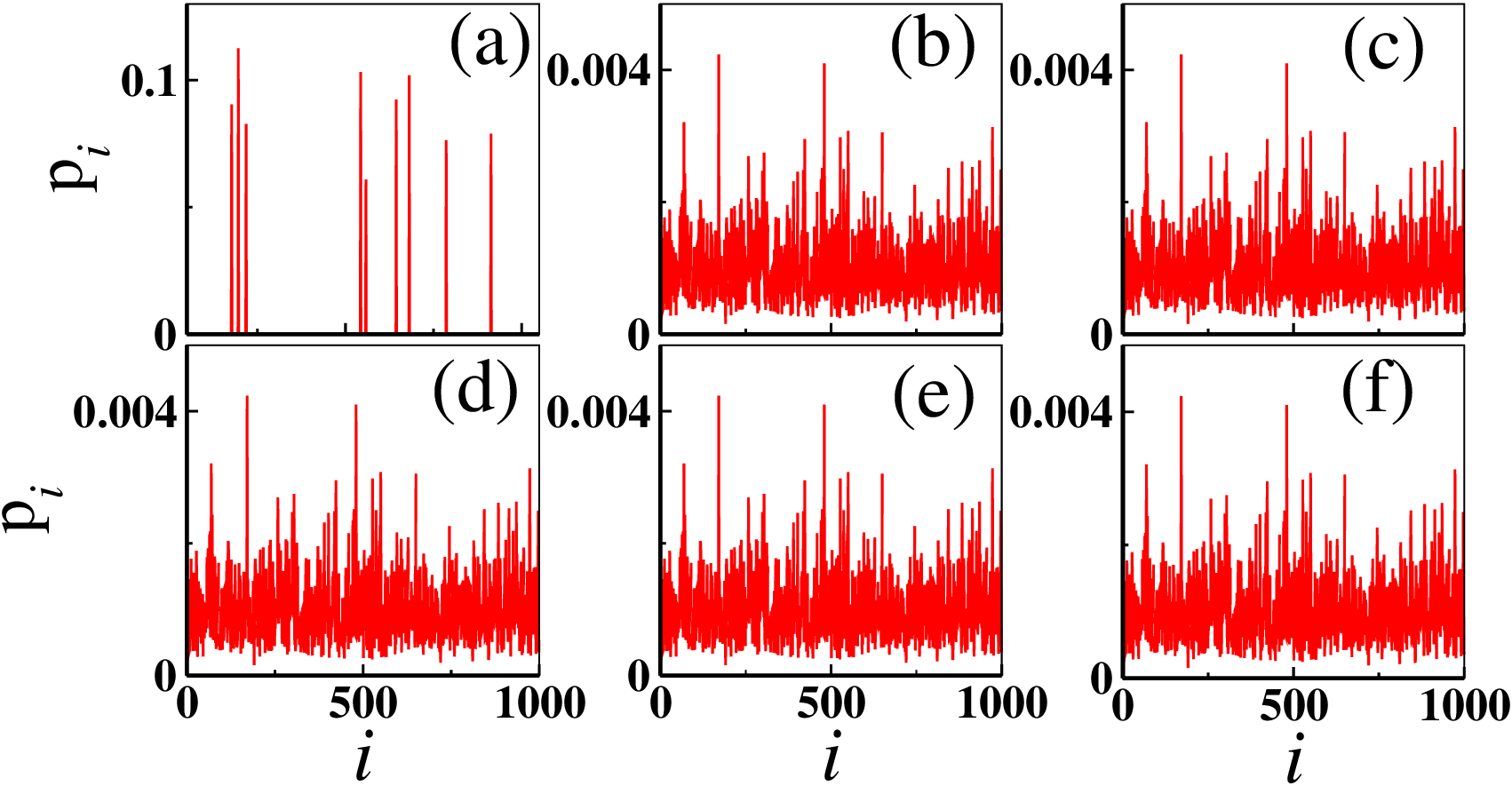} 
	\caption{Probability of finding the walker on different nodes for small-world networks with $p_{r} = 1$. (a) $t = 1$, (b) $t = 100$, and (c) $t = 1000$, (d) $t = 2000$ (e) $t = 5000$ (f) $t = 10000$;
		$N = 1000$ and $\langle k \rangle = 8$}
	\label{S4}
\end{figure}

Fig.~\ref{S4} demonstrate the evolution of probability distribution of finding the  walker on different nodes for  small-world networks with $p_{r} = 1$. It is evident from the figure that the walker reached the steady state even faster than $p_{r} = 0.1$ and also the probability distribution is homogeneous which makes the system more difficult to get localized.\\

